\documentclass[%
reprint,
nofootinbib,
amsmath,amssymb,
aps,
prl,
longbibliography
]{revtex4-1}

\usepackage{dsfont}
\usepackage{bbold}
\usepackage{mathrsfs}
\usepackage{amsmath}

\usepackage{graphicx}
\usepackage{dcolumn}
\usepackage{bm}

\usepackage{hyperref}
\newcounter{myequation}
\makeatletter
\@addtoreset{equation}{myequation}
\makeatother

\DeclareMathOperator{\tr}{\mbox{tr}}

\usepackage{color}
\definecolor{BV}{rgb}{0.1,0.,0.6}
\definecolor{R}{rgb}{0.9,0,0}
\definecolor{G}{rgb}{0.2,0.8,0.2}
\usepackage{comment}

\newcommand{\D}{{\mathcal D}}

\renewcommand{\L}{{\mathcal L}}

\newcommand{\be}{\begin{equation}}
	\newcommand{\ee}{\end{equation}}

\begin{document}

	\title{
	Duality between open systems and  closed bilayer systems,\\
	and
	thermofield double states as
	quantum many-body scars \\
}

\author{ Alexander Teretenkov$^1$}

%
\author{Oleg Lychkovskiy$^{2,1,3}$}

\affiliation{
	$^1$ Department of Mathematical Methods for Quantum Technologies,\\
	Steklov Mathematical Institute of Russian Academy of Sciences\\
	8 Gubkina St., Moscow 119991, Russia
}

\affiliation{%
	$^2$ Skolkovo Institute of Science and Technology\\
	Bolshoy Boulevard 30, bld. 1, Moscow 121205, Russia
}%

\affiliation{
	$^3$
	Russian Quantum Center, \\
	Bolshoy Boulevard 30, bld. 1, Moscow 121205, Russia
}

\date{\today}

\begin{abstract}
	We establish a duality between open many-body systems governed by the Gorini-Kossakowski-Sudarshan-Lindblad (GKSL) equation and satisfying the detailed balance condition on the one side, and closed bilayer systems with a self-adjoint Hamiltonian on the other side. Under this duality, the identity operator on the open system side maps to a  quantum many-body scar  of the dual Hamiltonian $\mathcal H$. This scar eigenstate has a form of a thermofield double state for a single-body conserved quantity entering the detailed balance conditions.  A remarkable feature of this thermofield scar is a tunable single-layer entanglement entropy controlled by the reservoir temperature on the open system side. Further, we identify broad classes of many-body open systems with nontrivial explicit eigen  operators $Q$ of the  Lindbladian superoperator.  The expectation values of the corresponding observables exhibit a simple exponential decay, $\langle Q\rangle_t=e^{-\Gamma t} \langle Q \rangle_0$, irrespectively of the initial state.  Under the above duality, these eigen operators give rise to additional (towers of) scars. Finally, we point out that more general superoperators (not necessarily of the GKSL form) can be mapped to self-adjoint Hamiltonians of bilayer systems harbouring scars, and provide an example thereof.
\end{abstract}


\maketitle

\noindent{\it Introduction.}~	
Dualities are precious gems of Physics. A duality provides a fortunate opportunity to look at the same phenomenon from different angles or, alternatively, to relate two seemingly different phenomena. Of particular value are instances when some object looks quite complex on the one side of the duality but turns out to be very simple on the other side. Then the duality becomes a powerful  tool for describing this object.

Here we establish a duality between open systems described by the GKSL equation and closed systems with doubled degrees of freedom (referred to as bilayer systems throughout the paper). Dualities of this type are  well known \cite{baranger1958problem, havel2003robust, Medvedyeva_2016_Exact} -- in a sense, their existence follows from the common mathematical nature of Lindbladian and Hamiltonian (both being linear (super-)operators) and a straightforward  counting of the respective configuration space dimensions  \cite[Chapter 4]{horn1994topics}. However, in general such dualities lead to non-Hermitian Hamiltonians. In contrast, the  duality presented here leads to  a self-adjoint Hamiltonian, provided the open system satisfies the detailed balance condition. This way the duality connects two theories with straightforward physical meanings.

We highlight the power of the duality by relating a simple object on the open system side -- the identity operator -- to a nontrivial object on the closed system side -- a quantum many-body scar.

Quantum many-body scars (further referred to as ``scars'' for brevity) are special, often explicitly known eigenstates of nonintegrable closed many-body systems \cite{Shiraishi_2017_Systematic,turner2018weak,serbyn2021quantum,moudgalya2022quantum}. They reside in the bulk of the spectrum and typically possess distinctly nonthermal properties. Scars can manifest themselves  in quantum quench  experiments through the lack of thermalization and persistent oscillations of local observables \cite{lin2017quasiparticle, bernien2017probing, chen2022error, su2023observation, zhang2023many}.

Remarkably, the scar obtained here turns out to be a {\it thermofield double state} with respect to a single-body observable which would be a conserved quantity in the absence of dissipation -- thus we adopt the term ``thermofield scar''.  The thermofield double state first emerged in thermal quantum field theory \cite{umezawa1982thermo} and further found applications in molecular dynamics \cite{Borrelli_2021_Finite}, black hole physics \cite{Maldacena_2003_Eternal,Maldacena_2013_Cool,hartman2013time} and quantum simulations \cite{Wu_2019_Variational, Zhu_2020_Generation,Francis_2021_Many-body}. Importantly, the {\it single-layer} bipartite entanglement entropy of the thermofield scar is thermal, with the temperature coinciding with the reservoir temperature on the open system side of the duality. In a special case of infinite temperature, the entanglement entropy is maximal and the thermofield scar reduces to the recently discovered Einstein-Podolsky-Rosen (EPR) scar \cite{langlett2022rainbow,wildeboer2022quantum}. At the same time, for other bipartitions the entanglement entropy of the thermofield scar is zero, which justifies the name ``scar'' \cite{serbyn2021quantum,moudgalya2022quantum}.

Further, we present broad families of open many-body systems where nontrivial eigen operators of the Lindbladian can be explicitly constructed. The duals of these eigen operators are additional scar states of the dual bilayer system.

Finally, we discuss a more general duality that connects self-adjoint Hamiltonians of bilayer systems to superoperators that are not of the GKSL form. We provide an example where such duality enables identification of nontrivial eigenstates of the bilayer Hamiltonian.

The rest of this paper is organized as follows. We start by introducing  the duality. Then we show that the duality implies the existence of the thermofield scar. After that we illustrate our general findings in a particular family of spin-$1/2$ models. Further, we demonstrate that in broad classes of dissipative models one can construct explicit  eigen operators of the adjoint Lindbladian, and that these operators map into additional scars in dual bilayer closed systems. Then we briefly discuss dualities between bilayer Hamiltonians and superoperators of non-GKSL form. We conclude by discussing our results in relation to other developments and outlining further prospects of our approach.


\medskip
\noindent{\it The duality.}~
On the open system side, we employ the GKSL equation in the Heisenberg representation \cite{Gorini_1976,Lindblad_1976},\cite[Sec. 3.2.3]{breuer2002theory}:
\begin{equation}\label{Lindblad equation_1}
	\partial_t O_t=\L^\dagger O_t\equiv i[H,O_t]+{\cal D}^\dagger O_t.
\end{equation}
Here  $O_t$ is a Heisenberg operator of some observable,  $\L$ is a Lindbladian superoperator (Lindbladian), $H$ is the Hamiltonian which is responsible for the coherent part of evolution, and $\cal D$ is the dissipation superoperator (dissipator) to be defined later. The operator $O_t$ acts in a Hilbert space $\mathscr{H}$. The expectation value $\langle O \rangle_t$ of the observable $O$ evolves in time according to  $\langle O \rangle_t=\tr \rho_0 \, O_t$, where $\rho_0$ is the initial state of the open system.

It is well-known that an open system can be formally mapped to a closed system with doubled degrees freedom, i.e.  with the Hilbert  space  $\mathscr{H}\otimes \mathscr{H}$. Under this map, any operator $O$ on the open system side maps to a vector $|O\rangle\rangle$ in $\mathscr{H}\otimes \mathscr{H}$,  the Lindbladian superoperator $i\L^\dagger$ maps to the ``Hamiltonian'' operator $\cal H$ acting in $\mathscr{H}\otimes \mathscr{H}$, and the GKSL equation \eqref{Lindblad equation_1} maps to the ``Schr\"odinger equation'' $i\partial_t |O_t\rangle\rangle={\cal H} |O_t\rangle\rangle$ \cite{baranger1958problem, havel2003robust, Medvedyeva_2016_Exact}. This map is merely a homomorphism between two linear spaces, the space of operators on the open system side and  the space of wave vectors on the closed system side. Importantly, under this map the operator $\cal H$ is never self-adjoint (except a trivial case of no dissipation, ${\cal D}=0$). While such non-Hermitian Hamiltonians can be useful  \cite{roccati2022non}, in general they do not correspond to any physical system.

\begin{figure}[t] 
	\centering
	\includegraphics[width=\linewidth]{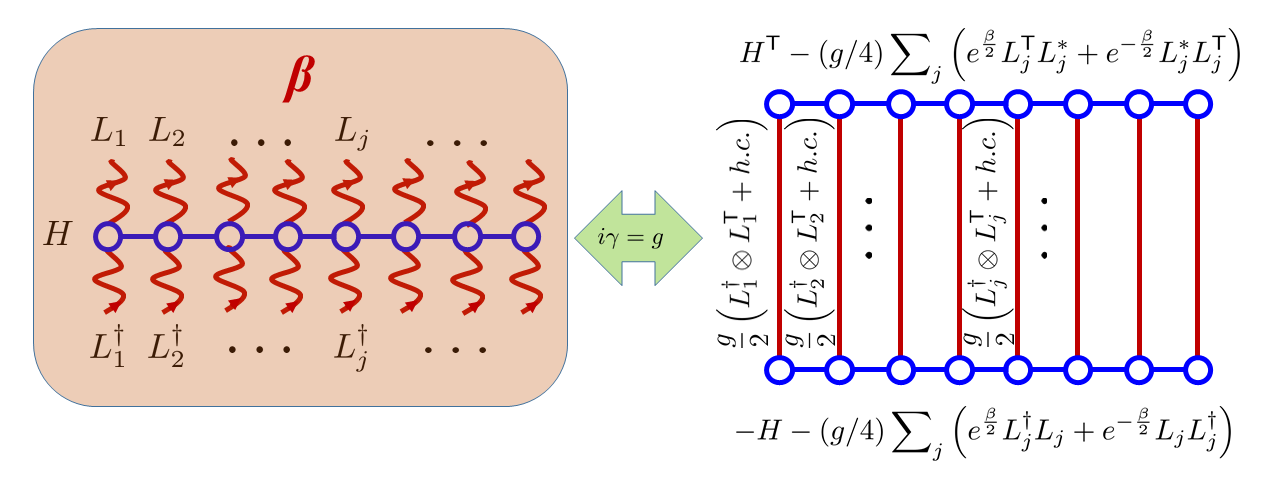}
	\caption{Illustration of the duality between an open system and a closed bilayer system. On the left an open system with a Hamiltonian $H$ and Lindblad operators $L_j$, $L_j^\dagger$ is schematically shown. The open system satisfies the detailed balance condition with the inverse temperature $\beta$, see eqs. \eqref{finiteTemperature},\eqref{annihilationLike}. On the right the dual closed system with doubled degrees of freedom is depicted. The Hamiltonian of the bilayer system is constructed  from the Hamiltonian of the open system and the Lindblad operators, see eq. \eqref{finiteTemperatureHamiltonian}. }
	\label{fig}
\end{figure}

Our goal is to construct a {\it duality} that will map $i\L^\dagger$ to a self-adjoint Hamiltonian.  To this end we employ a  Wick rotation of the dissipative coupling constant  on top of the linear homomorphism.  We will demonstrate that one can obtain  a self-adjoint Hamiltonian this way, provided the dissipator satisfies the {\it detailed balance condition}.

First we recall the general form of the GKSL dissipator, $\D^\dagger O_t =  \sum_j \left(\tilde L_j^{\dagger}  O_t \tilde L_j - (1/2)\{\tilde L_j^{\dagger}\tilde L_j,  O_t \}\right)$, where $\{\bullet,\bullet\}$ is an anticommutator and $\tilde L_j$ are arbitrary  operators \cite{Gorini_1976,Lindblad_1976,breuer2002theory}. The dissipator is said to satisfy the detailed balance condition if it can be written in the form \cite[Sec. 1.3.4]{Alicki_2007_Quantum}
\begin{align}
	\D^\dagger O_t = \frac{\gamma}{2} \sum_j  \biggl(  &e^{\frac{\beta}{2}}\left( L_j^\dagger O_t L_j-\frac12 \{L_j^\dagger  L_j,O_t\}\right) \nonumber\\
	+& e^{-\frac{\beta}{2}}\left( L_j O_t L_j^\dagger-\frac12 \{L_j  L_j^\dagger,O_t\} \right) \biggr), \label{finiteTemperature}
\end{align}
where  $ \beta $ is a real parameter having the meaning of the inverse reservoir temperature in microscopic derivations of the GKSL equation \cite[Sec. 3.3.2]{breuer2002theory}, $\gamma>0$ is the dissipative coupling constant  and (rescaled) Lindblad operators $L_j$ are annihilation-like operators with respect to some $H_0$ that commutes with $H$, i.e.
\begin{equation}\label{annihilationLike}
	[\beta  H_0, L_j] = - \beta L_j, \quad [H_0,H]=0.
\end{equation}
The dissipator \eqref{finiteTemperature} can be written in  the standard GKSL form by replacing rescaled Lindblad operators $L_j$ by conventional Lindblad operators $\tilde L _j$ according to
$\tilde L_{2j}= \sqrt{\gamma} e^{\frac{\beta}{4}} L_j$ , $\tilde L_{2j + 1}= \sqrt{\gamma} e^{-\frac{\beta}{4}} L_j^{\dagger}$.
Note that the detailed balance condition entails the constraint $\tilde L_{2j}^\dagger=e^{\frac{\beta}{2}} \tilde L_{2j + 1}$.

The detailed balance condition \eqref{finiteTemperature}, \eqref{annihilationLike}  naturally emerges when deriving the GKSL equation for a system interacting with a thermal reservoir in weak coupling or low density limits \cite[Sec. 1.3.4]{Alicki_2007_Quantum}.\footnote{In fact, the detailed balance condition has a more general form, which can be  obtained by replacing $\beta \rightarrow \beta \omega_j$ in the right hand sides of eqs. \eqref{finiteTemperature} and \eqref{annihilationLike}, where $\omega_j$ can be all different \cite[Sec. 1.3.4]{Alicki_2007_Quantum}.  Further generalizations of the  detailed balance in quantum systems are discussed in \cite{Fagnola_2010_Quantum}. While our construction can be generalized appropriately, we leave such generalizations to further studies.  } Under this condition, the GKSL equation entails a classical Markov equation with classical detailed balance for populations of the conserved quantity  $H_0$  (see e.g. \cite[Section 3.3.2]{breuer2002theory}).

Note that $\beta$ is not cancelled in eq. \eqref{annihilationLike} for a purpose: at infinite temperature, $\beta=0$, this equation is satisfied for arbitrary $L_j$. Importantly, in this case the Lindblad operators can be redefined to be self-adjoint, $L_j'=(L_j+L_j^\dagger)/\sqrt2$, $L_j''=i\,(L_j-L_j^\dagger)/\sqrt2$, with the dissipator reducing to
$
\D^\dagger O_t=-\frac\gamma2\,\sum_j\big([L'_j,[L_j',O_t]]+[L_j'',[L_j'',O_t]]\big).
$
Vice versa, if one takes all $L_j$ in eq. \eqref{finiteTemperature} to be self-adjoint from the outset, the dissipator assumes the same form,
\begin{equation}\label{infiniteTemperatrueDissipator}
	\D^\dagger O_t=-\frac{\gamma'}2\,\sum_j[L_j,[L_j,O_t]],\quad L_j^\dagger=L_j,
\end{equation}
where $\gamma'=\gamma \cosh(\beta/2)$. We will refer to the dissipators of  the form \eqref{infiniteTemperatrueDissipator} as infinite-temperature dissipators, although they  can also arise in other settings, e.g. in the singular-coupling limit \cite[Sec. 3.3.3]{breuer2002theory}.

To define the duality, we first fix an orthonormal basis $\{|n\rangle\}$ in the Hilbert space $\mathscr{H}$. After that we map operators $O$ on the open system side to vectors $ |O\rangle\rangle_{\beta}$ on the closed system side according to
\begin{align}
	O=&\sum_{m,n}O_{mn} |m\rangle\langle n| ~ \longleftrightarrow~ \nonumber\\ &|O\rangle\rangle_{\beta}=\sum_{m,n}O_{mn}  e^{- \frac{\beta}{4} H_0} |m\rangle \otimes  e^{-\frac{\beta}{4} H_0^T} | n \rangle, \label{operatorVectorCorrespondenceFiniteTemperature}
\end{align}
where the subscript $\beta$ emphasizes that the dual vector depends on the inverse temperature.
This map induces the map between superoperators acting on $O$ and operators acting on $|O\rangle\rangle_{\beta}$. Applied to the Lindbladian, it reads $i  \, \L^\dagger \,\longleftrightarrow\, \mathcal{H}_{\beta}$  with \cite{supp}
\begin{align}
	\mathcal{H}_{\beta}=&\left(-H-\frac{g}4 \sum_j\left( e^{ \frac{\beta}{2}}L_j^\dagger L_j +  e^{- \frac{\beta}{2}} L_j L_j^\dagger\right) \right)\otimes \mathbb{1}\nonumber\\[0.8em]
	& +\mathbb{1}\otimes \left(H^T-\frac{g}4 \sum_j \left(e^{ \frac{\beta}{2}} L_j^TL_j^* + e^{ -\frac{\beta}{2}} L_j^*L_j^T\right)\right) \nonumber\\[0.8em]
	&+\frac{g}{2}\sum_j  \biggl(L_j^\dagger\otimes L_j^T + L_j\otimes L_j^{*}\biggr). \label{finiteTemperatureHamiltonian}
\end{align}
Here the Wick rotation $g = i\gamma $ of the dissipative coupling constant has been performed. One can check directly that whenever $g$ is real, the Hamiltonian ${\cal H}_\beta$ is self-adjoint, as desired.

The Hamiltonian ${\cal H}_\beta$ is naturally interpreted as the Hamiltonian of a bilayer system:  the first (the second) line of eq. \eqref{finiteTemperatureHamiltonian} represents the Hamiltonian of the first (the second) layer, while the third line -- the interaction between the layers, see Fig. \ref{fig} for illustration. The layers can have a direct geometrical meaning   \cite{barkeshli2014synthetic, Hubert_2019_Attractive,gall2021competing,bohrdt2022strong} or, alternatively, represent  a binary  internal degree of freedom, such as fermionic spin in the Hubbard model \cite{supp}.

It should be stressed that the Wick rotation along, without the similarity transformation \eqref{operatorVectorCorrespondenceFiniteTemperature}, does not lead to a self-adjoint dual Hamiltonian unless all Lindblad operators can be chosen to be self-adjoint, as in eq. \eqref{infiniteTemperatrueDissipator}. The latter case was considered in our previous paper \cite{teretenkov2024exact}. Furthermore,  a {\it general} Lindbladian can not be mapped to a self-adjoint Hamiltonian by means of any similarity transformation and/or Wick rotation. This can be seen from the fact that such mapping would imply the diagonalizability of the Lindbladian, which is not guaranteed even for a two-level system \cite{Lendi_1987_Evolution}. These observations highlight the crucial role of the detailed balance condition for obtaining the duality.

Note that a different choice of the basis in eq. \eqref{operatorVectorCorrespondenceFiniteTemperature} leads to a different duality. In particular,  matrix transposition and complex conjugation are basis-dependent operations, therefore $H^T$, $L_j^T$ and $L_j^{*}$ entering eq. \eqref{finiteTemperatureHamiltonian} depend on the basis choice. One consequence of the freedom to chose the basis is discussed in the next section.

The duality \eqref{operatorVectorCorrespondenceFiniteTemperature}, \eqref{finiteTemperatureHamiltonian} is the first main result of the present Letter.


\medskip
\noindent{\it The thermofield scar.}~
The GKSL equation in the Heisenberg picture \eqref{Lindblad equation_1} always preserves identity operator $\mathbb{1}$, i.e.
\begin{equation}\label{identity as an eigenoperator of Lindbladian}
	i\L^\dagger \, \mathbb{1} =0.
\end{equation}
The duality then implies that the dual vector,
\begin{equation}\label{thermofield double state}
	|\mathbb{1} \rangle\rangle_\beta=\sum_n  e^{- \frac{\beta}{4} H_0} |n \rangle \otimes  e^{-\frac{\beta}{4} H_0^T} | n \rangle,
\end{equation}
is an (unnormalized) eigenstate of the dual Hamiltonian,
\begin{equation}
	{\mathcal H}_\beta |\mathbb{1} \rangle\rangle_\beta = 0.
\end{equation}
This is the second main result of the Letter.

\begin{figure}[t] 
	\centering
	\includegraphics[width=\linewidth]{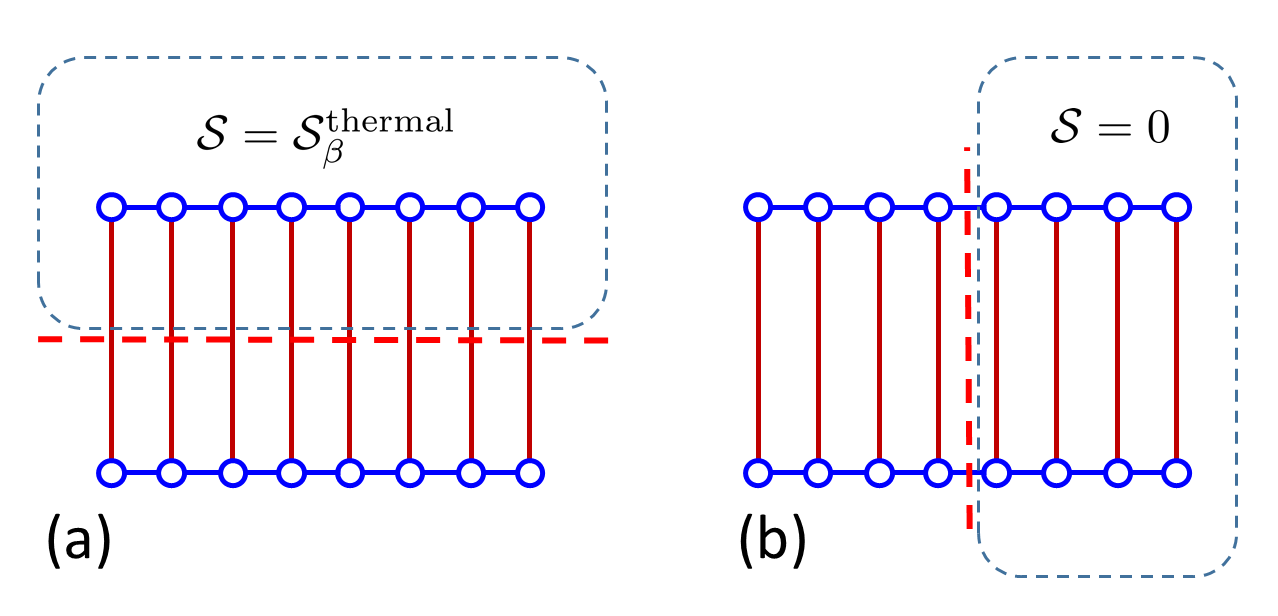}
	\caption{Entanglement of the thermofield scar \eqref{thermofield double state} with respect to different bipartitions of a bilayer system. (a) For the bipartition that singles out one layer and traces out another one, the entanglement entropy equals the thermal entanglement entropy for the Hamiltonian~$H_0$ at  temperature~$\beta$. (b)~The entanglement entropy vanishes   for a transverse bipartition that does not cut the interlayer bonds (here the underlying basis vectors are assumed to be of product form, see the text for details).}
	\label{fig 2}
\end{figure}

The state \eqref{thermofield double state} is a thermofield double state~\cite{umezawa1982thermo} for $H_0$. It possesses a remarkable entanglement structure highly dependent on the bipartition. If one traces out one layer of the thermofield scar~\eqref{thermofield double state}, see Fig. \ref{fig 2} (a), one obtains a reduced state of the second layer which turns out to be  a Gibbs state with respect to  $H_0$  \cite[Section 32.5]{nastase2023quantum}, \cite[Eq. \eqref{thermofieldTrace}]{supp},
\begin{equation}\label{GibbsState}
	\operatorname{Tr}_{\text{one layer}}|\mathbb{1}\rangle\rangle_{\beta} \langle \langle\mathbb{1} |_{\beta} =  e^{- \beta H_0},
\end{equation}
where normalization is omitted.\footnote{Curiously,  the  reduced state \eqref{GibbsState} is a steady state of the Lindbladian $\L$.} As a consequence, the entanglement entropy between layers is merely the thermal equilibrium entropy for a system with the Hamiltonian $H_0$ and inverse temperature $\beta$.

However, for ``transverse'' bipartitions shown in Fig~\ref{fig 2}(b), the picture is completely different. The precise value of the entanglement entropy for such bipartitions depends on the choice of the basis underlying the duality. In the extreme case where each basis vector is of product form, the thermofield state \eqref{thermofield double state} factorizes for any transverse bipartition and the entanglement entropy is equal to zero. This is explicitly demonstrated for an exemplary system considered in what follows.

In general, the double layer Hamiltonian \eqref{finiteTemperatureHamiltonian} is nonintegrable  and non-localized (here we assume no disorder on the open system side).  Further, zero energy is, in general, in the middle of the spectrum. Therefore the zero energy eigenstate \eqref{thermofield double state} possessing  a zero entanglement entropy for certain bipartitions is classified as a scar~\cite{serbyn2021quantum,moudgalya2022quantum}. We adopt the term ``thermofield scar'' for this state.

We remark that in some cases, even after the thermofield scar and related towers of special eigenstates discussed below are removed from the spectrum, the rest of the spectrum may not appear completely chaotic for various reasons. In this case the term ``scar'' applied to the eigenstate \eqref{thermofield double state} may become a misnomer, since scars are supposed to be embedded in the surrounding of chaotic eigenstates. With this caveat in mind, we nevertheless keep using the term ``thermofield scar'' for the sake of brevity.

A particular case of the thermofield scar \eqref{thermofield double state} with $\beta=0$ has been recently  discovered in refs. \cite{langlett2022rainbow,wildeboer2022quantum} (without using or recognizing the duality).\footnote{See also ref.  \cite{cottrell2019build} where  a local bilayer Hamiltonian is constructed whose ground state is approximately  the thermofield double state. } It has been tagged as the EPR scar in ref. \cite{wildeboer2022quantum}, for reasons that will be clarified in the example below. Refs. \cite{langlett2022rainbow,wildeboer2022quantum} contain an in-depth analysis of the various features of infinite-temperature thermofield scars, including a numerical demonstration of persistent oscillations of local observables for a quantum quench.




\medskip
\noindent{\it An example.}~
Consider an arbitrary-range isotropic  dissipative  model of $N$ spins $1/2$ with the  Hamiltonian
\begin{equation}\label{H exemplary XY}
	H= \sum_{i<j} w_{ij} \left( \sigma^x_i \sigma^x_{j}+   \sigma^y_i \sigma^y_{j} \right),
\end{equation}
$w_{ij}$ being arbitrary real coupling constants, and the dissipator of the form \eqref{finiteTemperature} with
\begin{equation}\label{L exemplary XY}
	L_j=\sigma^-_j\equiv (1/2)(\sigma^x_j- i\sigma^y_j).
\end{equation}
This model satisfies the detailed balance condition \eqref{annihilationLike} with respect to the total magnetization in the $z$-direction,
\begin{equation}
	H_0 = S^z= \frac12 \sum_{j} \sigma_{j}^z,\quad [H_0,H]=0.
\end{equation}

Fixing the basis to be the common eigenbasis of all $\sigma^z_j$, one obtains from eq. \eqref{finiteTemperatureHamiltonian} the explicit dual self-adjoint Hamiltonian:
\begin{align}
	\mathcal{H}_{\beta} =& -  \sum_{i<j}  w_{ij}  \left(\,\sigma^x_i \sigma^x_{j}+  \, \sigma^y_i \sigma^y_{j} \right) - \frac{g}{4} \sinh \frac{\beta}{2} \sum_j \sigma^z_j \nonumber\\
	& + \sum_{i<j} w_{ij} \left(  \,\tau^x_i \tau^x_{j}+  \, \tau^y_i \tau^y_{j} \right) - \frac{g}{4} \sinh \frac{\beta}{2} \sum_j \tau^z_j\nonumber\\
	&+\frac{g}{4}\sum_j  \bigl(\sigma^x_j \tau^x_j - \sigma^y_j \tau^y_j \bigr) - N \frac{g}{2}  \cosh \frac{\beta}{2}.	
	\label{dual sigma pm example}
\end{align}
Here we have changed the notations as  $\sigma^{x,y,z}_{j} \otimes \mathbb{1} \rightarrow \sigma^{x,y,z}_{j}$ and  $ \mathbb{1} \otimes \sigma^{x,y,z}_{j}\rightarrow \tau^{x,y,z}_{j}$, so that now   $\sigma^{x,y,z}_{j}$ refers  to the first layer while $\tau^{x,y,z}_{j}$ -- to the second one. The first (the second) line in eq. \eqref{dual sigma pm example} is the Hamiltonian of the first (second) layer and the third line is the interlayer interaction plus an additional constant term.

The thermofield scar \eqref{thermofield double state} in  the considered model can be presented in the factorized form,
\begin{equation}\label{thermofield double state example}
	|\mathbb{1} \rangle\rangle_\beta=\bigotimes_{j=1}^N \Big(e^{-\frac{\beta}{4}}|\uparrow\,\uparrow\rangle_j+e^{\frac{\beta}{4}}|\downarrow\,\downarrow\rangle_j\Big),
\end{equation}
where each factor is a state of two adjacent spins residing in different layers.
At $\beta=0$ each factor represents a EPR pair, thus the term ``EPR scar'' in ref. \cite{wildeboer2022quantum}.


\begin{figure}[t] 
	\centering
    \includegraphics[width=\linewidth]{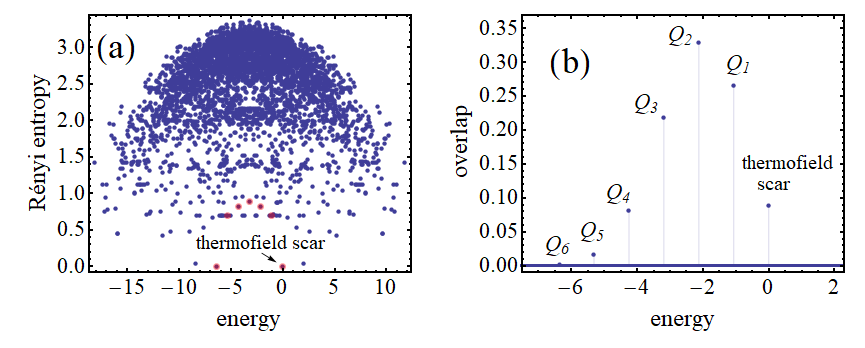}\\
	\includegraphics[width=0.9\linewidth]{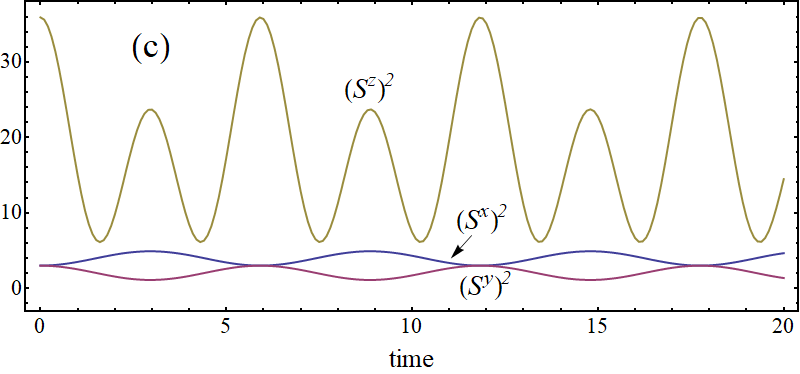}
	\caption{Scars and scarred dynamics for the bilayer system~\eqref{dual sigma pm example} with $N=6$ spins in each layer. The Hamiltonian parameters are $g=1$, $\beta=0.7$, $\omega_{j\,j+1}=1$, $\omega_{j\,j+2}=0.5$ and $\omega_{ij}=0$ otherwise. (a)~R\'enyi entropy~$(-\log \tr \rho_E^2)$ of eigenstates with respect to the transverse bipartition (see Fig.~\ref{fig 2}(b)) in two equal parts ($\rho_E$ is the corresponding reduced density matrix of an eigenstate). Highlighted are the thermofield scar~$|\mathbb{1} \rangle\rangle_\beta$ and the scars $|Q_n \rangle\rangle_\beta$, $n=1,2,\dots,6$. (b)~ Overlaps~$|\langle   E|\Psi_0\rangle|^2$ between the eigenstates and the product initial state  $\Psi_0=|\downarrow\downarrow\dots\downarrow\rangle$ (all spins down in both layers). One can see that this initial state is expanded over the thermofield scar and the scars~$|Q_n \rangle\rangle_\beta$. (c)~Persistent oscillations of the squared total spin polarizations $\left(S^{x,y,z}\right)^2$ after initialization in the state~$\Psi_0$.
	\label{fig scars}
}
\end{figure}

\medskip
\noindent{\it Explicit eigen operators of the Lindbladian.}~
An adjoint Lindbladian $\L^\dagger$ can have other eigen operators apart from the trivial identity operator considered above. Here we focus on eigen operators $Q$ that would be integrals of motion without dissipation,
\begin{equation}\label{HIoM}
	[H,Q]=0.
\end{equation}
They are referred to as  Hamiltonian integrals of motion (HIoM). We assume that a HIoM  $Q$ is an eigen operator of the adjoint dissipator and hence of the adjoint Lindbladian,
\begin{equation}\label{condition for Q}
	\D^{\dagger} Q =\L^{\dagger} Q =-\Gamma Q
\end{equation}
where $\Gamma$ is a real non-negative number. We refer to such HIoMs as eigen-HIoMs.  The GKSL equation \eqref{Lindblad equation_1} is immediately solved for an eigen-HIoM $Q$ with the result
\begin{equation}\label{dynamics of Q}
	Q_t =e^{-\Gamma t} Q.
\end{equation}
The expectation value of the eigen-HIoM experiences the same exponential decay  for an {\it arbitrary} initial condition,
$
\langle Q \rangle_t =e^{-\Gamma t} \langle Q \rangle_0.
$
Note that  in the case of integrable $H$ a useful generalization of this construction exists \cite{supp}.

Remarkably, there are  broad classes of models where  eigen-HIoMs can be explicitly found. The model \eqref{H exemplary XY}, \eqref{L exemplary XY} is one example. There the condition \eqref{condition for Q} is satisfied for  a set $Q_n$ of eigen-HIoMs that are  polynomials  constructed from  $S^z$, $n\leq N$ being the polynomial degree. The first two of them read \cite{supp}
\begin{align}\label{HIoMs exemplary}
	Q_1 & = S^z + \frac{N}{2}\tanh\frac{\beta}2 \cdot {\mathbb 1},            & \Gamma_1 & = \gamma\cosh\frac{\beta}2,  \nonumber \\
	Q_2 & = \left(S^z\right)^2 +(N-1)\tanh \frac{\beta}{2}  S^z \nonumber \\
	&+\frac{N}{4}\left((N-1) \tanh^2 \frac{\beta}{2} - 1\right)\cdot {\mathbb 1},        &
	\Gamma_2 & = 2 \, \gamma\cosh\frac{\beta}2, \nonumber   \\
	\dots
\end{align}
In addition, the single-layer Hamiltonian $H$ is also an eigen-HIoM with  $\Gamma=\gamma \cosh\frac{\beta}2$.

Other families of models with eigen-HIoMs are presented in the Supplement~\cite{supp}. They include spin and fermionic models with various dissipative terms.

\medskip
\noindent{\it Dual scars.}~
Explicit (towers of) eigen-HIoMs $Q$ of the adjoint Lindbladian discussed in the previous section map to  (towers of)  scars $|Q\rangle\rangle_\beta$ under the duality according to eq.~\eqref{operatorVectorCorrespondenceFiniteTemperature}.  Remarkably, these scars can be obtained from the thermofield scar as
\begin{equation}\label{additional scars}
	|Q\rangle\rangle_\beta = Q(\beta)    \otimes \mathbb{1} \; | \mathbb{1}\rangle \rangle_\beta =   \mathbb{1} \otimes Q^T(-\beta)   \;  | \mathbb{1}\rangle \rangle_\beta,
\end{equation}
where  $ Q(\beta) =   e^{- \frac{\beta}{4} H_0} Q  e^{\frac{\beta}{4} H_0}$, which can be verified by a direct calculation \cite[Eq. \eqref{finiteTemperatureInTermsOfScar}]{supp}. An infinite-temperature analog of this formula was presented in refs. \cite{langlett2022rainbow,wildeboer2022quantum}. 

Explicit expressions for scars of the model  \eqref{dual sigma pm example} dual to eigen-HIoMs \eqref{HIoMs exemplary}  can be found in the Supplement~\cite{supp}. Note that they do not have a strictly vanishing entanglement entropy for the transverse bipartition.  In Fig. \ref{fig scars} we show these scars immersed in the spectrum of the model  \eqref{dual sigma pm example}. We also demonstrate that, analogously to the infinite-temperature case \cite{langlett2022rainbow}, the initial product state $\Psi_0=|\downarrow\downarrow\dots\downarrow\rangle$ (all spins down in both layers) has large overlaps with the thermofield scar \eqref{thermofield double state} and the scars from the tower dual to HIoMs \eqref{HIoMs exemplary}, which entails persistent oscillations of the total spin  projections squared, $\left(S^\alpha\right)^2(t)$, $\alpha=x,y,z$.

Curiously, the spectrum of the model \eqref{dual sigma pm example} apparently contains other scars. Two of them, $|P_1\rangle=\bigotimes_{j=1}^N |\uparrow\,\downarrow\rangle_j$ and  $|P_2\rangle=\bigotimes_{j=1}^N |\downarrow,\uparrow\rangle_j$ (spins up in one layer, spins down in another layer), are of the product form and do not depend on $\beta$. While they can be understood in terms of the duality \cite{supp}, in fact they are examples of curious special eigenstates generic for a wider class of bilayer systems beyond the duality \cite{supp}. We leave deeper exploration of the model \eqref{dual sigma pm example} for further studies.

The observation that numerous families of dissipative many-body models possess (towers of) explicit isolated eigen operators of the adjoint Lindbladian, and that these eigen operators turn into (towers of) quantum many-body scars in dual bilayer systems constitute the third main result of the present paper.



\medskip
\noindent{\it Beyond GKSL.}~ Finally, let us briefly comment on dualities in a more general perspective. The dimension counting  tells us that, in fact, an arbitrary self-adjoint bilayer Hamiltonian $\mathcal{H}$ can be mapped to some superoperator $\mathcal S$ acting on operators in a Hilbert space corresponding to a single layer. This superoperator is not necessarily of the GKSL form, therefore, at the first sight, such a duality misses much of its physical appeal.\footnote{Note that non-Markovian dynamics of open  quantum systems can lead to GKSL-like, but non-GKSL generators \cite{hall2014canonical}.} However, it can still be very useful, in particular for identifying special eigenstates of $\mathcal{H}$. To illustrate the latter claim, consider a bilayer spin-1/2 Hamiltonian $\mathcal{H}=\sum_{i<j}  w_{ij}  \left(\,\sigma^x_i \sigma^z_j +\tau^x_i\tau^z_j\right)+ \sum_{j} \sigma^y_j\tau^y_j$ (notations are the same as in the Hamiltonian  \eqref{dual sigma pm example}). The infinite-temperature version of the map \eqref{operatorVectorCorrespondenceFiniteTemperature} leads to the  superoperator  ${\mathcal S} O =\{H,O\}-  \sum_{j} \sigma^y_j\, O \, \sigma^y_j$, where $H=\sum_{i<j}  w_{ij}  \sigma^x_i \sigma^z_j $  and $\{\bullet,\bullet\}$ is the anticommutator. Clearly, ${\mathcal S}$ is not of the GKSL form. However, one easily identifies two eigen operators $Q_{x,z}$ of ${\mathcal S}$,  $Q_{x,z}=\prod_j \sigma_j^{x,z}$, satisfying   ${\mathcal S} \, Q_{x,z} = N\, Q_{x,z}$. They correspond to two eigenstates $|Q_x \rangle\rangle=\bigotimes_j \left( |\uparrow\,\downarrow\rangle_j+|\downarrow\,\uparrow\rangle_j \right)$ and $|Q_z \rangle\rangle=\bigotimes_j \left( |\uparrow\,\uparrow\rangle_j-|\downarrow\,\downarrow\rangle_j \right)$  of the bilayer Hamiltonian $\mathcal{H}$ (notations are the same as in eq. \eqref{thermofield double state example}). This example reveals that viewing bilayer systems through the prism of a general duality of type \eqref{operatorVectorCorrespondenceFiniteTemperature} can help identify their features concealed otherwise.  Exploring such dualities constitutes a promising direction for further research.


\medskip
\noindent{\it Discussion and outlook.}~
Eigen operators of the Lindbladian dual to scars in bilayer closed systems
can be thought of as ``operator scars'' in open systems.\footnote{See refs. \cite{Wang_2024_Embedding,deLeeuw2023hidden} for somewhat different perspectives on operator scars.} Their pure exponential decay is in sharp contrast to the ergodic evolution of a typical operator that explores the whole operator space allowed by symmetries \cite{viswanath2008recursion,Parker_2019}. From a different angle, the pure exponential decay of HIoMs resembles purely oscillating behavior of  decoherence free states \cite{lidar1998decoherence, agredo2014decoherence} and subalgebras \cite{agredo2022decoherence}. It can be interesting to look for a deeper connection between these phenomena. For the moment we just remark that  the decoherence free algebras in open systems are dual to the states of closed bilayer systems with the decoupled dynamics of the layers.

A phenomenon closely related to quantum many-body scarring is Hilbert space fragmentation \cite{Znidaric_2013_Coexistence,Sala_2020_Ergodicity,Khemani_2020_Localization,Moudgalya_2021_Thermalization}. As we discuss in the Supplement \cite{supp},  the  Hilbert space fragmentation in closed bilayer systems is dual to the operator space fragmentation  \cite{Essler_2020_Integrability} (see also \cite{Li_2023_Hilbert}) in open systems. The latter phenomenon typically occurs in open systems with integrable Hamiltonians \cite{Essler_2020_Integrability, Robertson_2021_Exact, Li_2023_Hilbert, Horstmann_2013_Noise-driven, Bakker_2020, Eisler_2011, Temme_2012, Shibata_2019_Dissipative, Dolgirev_2020_Non-Gaussian, Turkeshi_2021_Diffusion, haken1973exactly, Rahmani_2015_Dynamics, Foss-Feig_2017_Solvable, Budini_2021_Solvable, deLeeuw2023hidden, teretenkov2024exact}.
Importantly, the structure of the fragmentation is much more transparent on the open system side of the duality \cite{dodonov1988evolution, Prosen_2008_Third, teretenkov2019dynamics, teretenkov2020dynamics, nosal2020exact, ivanov2022dynamics, linowski2022dissipative, nosal2022higher, Barthel_2022,Zhang_2022, Zunkovic_2014_Closed}, analogously to the case of scars.

Much of the theory presented here remains valid in the case of time-dependent Hamiltonian and Lindblad operators. In particular, the thermofield state \eqref{thermofield double state} remains a zero energy stationary state of the now-time-dependent dual bilayer Hamiltonian in this case, provided the detailed balance conditions \eqref{finiteTemperature} are satisfied at each moment of time. This paves the way to Floquet scars \cite{Mizuta_2020_Exact,Yarloo_2020_Homogeneous,Mukherjee_2020_Collapse,moudgalya2022quantum} in periodically driven bilayer systems, as well as to even more exotic stationary states of bilayer Hamiltonians that depend on time aperiodically. We leave this direction of study for further work.

Much work has been made on identifying and characterizing nonequilibrium steady states of dissipative system  supporting a steady current \cite{Znidaric_2010_Exact,Znidaric_2011_Solvable,Prosen_2011_Exact,Eisler_2011,Temme_2012,Karevski_2013_Exact,Ilievski_2014_Quantum,
	Prosen_2015_Matrix,Popkov_2017_Solution,Ribeiro_2019_Integrable}. As a potentially interesting tweak, one can attempt to map such states to scars in dual bilayer systems. This might be possible, in particular, in the case of  Lindblad operators that form detailed balance pairs as in ref.~\cite{trushechkin2019general}.

It can be interesting to explore the freedom to use different bases for defining the duality is yet to be explored. Here we have focused on the simplest case of product state bases; more complex bases will lead to dualities with more complex properties and, in particular, to thermofield scars with more intricate entanglement structure.

Finally, we note that the reported duality might aid quantum simulations of open systems \cite{david2024faster} and shed new light on topological effects in open systems \cite{Niu_2024_Topological}.
	
\medskip
\begin{acknowledgments}
	\noindent{\it  Acknowledgments.}
	We thank I. Ermakov for useful discussions. The work of O. Lychkovskiy was supported by
	the Foundation for the Advancement of Theoretical Physics
	and Mathematics “BASIS” under Grant No. 22-1-2-55-1.
\end{acknowledgments}

	\bibliography{scars,operator_scar_misc,open_systems,recursion_method}
	

\clearpage

\setcounter{page}{1}

\onecolumngrid

\renewcommand{\theequation}{S\arabic{equation}}
\stepcounter{myequation}

\appendix

	\section*{Supplementary material for Letter\\
``Duality between open systems and  closed bilayer systems,\\
	and
	thermofield double states as
	quantum many-body scars'' \\
by Alexander Teretenkov and Oleg Lychkovskiy
	}

\section{S1. Basic formulae for dual vectors}

First of all let us recall the basic formulae for the infinite-temperature case. Remark that
\begin{equation}\label{EPRscar}
	|\mathbb{1}\rangle\rangle_{0} = \sum_{n}  |n\rangle \otimes  | n \rangle,
\end{equation}
then by definition \eqref{operatorVectorCorrespondenceFiniteTemperature} at $\beta = 0$ we have
\begin{align}
	|AB\rangle\rangle_{0}=\sum_{m,n}(AB)_{mn}  |m\rangle \otimes  | n \rangle =  \sum_{m,n} \langle m |AB  |n\rangle  |m\rangle \otimes  | n \rangle =   \sum_{m,n, k} \langle m |A |k\rangle\langle k|  B  |n\rangle  |m\rangle \otimes  | n \rangle \nonumber \\
	=   \sum_{m,n, k}   |m\rangle \langle m |A |k\rangle \otimes  | n \rangle \langle n|  B^T  |k\rangle = \sum_{ k}  A |k\rangle \otimes   B^T  |k\rangle  	= A \otimes B^T \; |\mathbb{1}\rangle\rangle_{0} \label{equivalentForms}
\end{align}
for arbitrary operators $A$ and $B$.
In particular, for an operator $O$ we have
\begin{equation}
	|O\rangle\rangle_{0}=	  O \otimes \mathbb{1} \; |\mathbb{1}\rangle\rangle_{0}  = \mathbb{1}\otimes O^T \; |\mathbb{1}\rangle\rangle_{0}. \label{equivalentFormsForSingleOperator}
\end{equation}

The dual vector in the finite-temperature case can be expressed in terms of one in the infinite-temperature case as
\begin{equation}
	|O\rangle\rangle_{\beta} = \sum_{m,n}O_{mn}  e^{- \frac{\beta}{4} H_0} |m\rangle \otimes  e^{-\frac{\beta}{4} H_0^T} | n \rangle = e^{- \frac{\beta}{4} H_0} \otimes  e^{-\frac{\beta}{4} H_0^T}  \; |O\rangle\rangle_0 = e^{- \frac{\beta}{4} (H_0 \otimes  \mathbb{1} + \mathbb{1} \otimes H_0^T)} |O\rangle \rangle_0. \label{relationsBetweenFiniteAndInfinite}
\end{equation}
Then using \eqref{equivalentFormsForSingleOperator} and \eqref{equivalentForms} we obtain
\begin{align}
	|O\rangle\rangle_{\beta} =( e^{- \frac{\beta}{4} H_0}  \otimes  e^{-\frac{\beta}{4} H_0^T})  (O \otimes \mathbb{1}) | \mathbb{1}\rangle\rangle_{0}  &=  e^{- \frac{\beta}{4} H_0} O \otimes  e^{-\frac{\beta}{4} H_0^T}  \; | \mathbb{1}\rangle\rangle_{0} \nonumber\\
	 &=e^{- \frac{\beta}{4} H_0} O  e^{-\frac{\beta}{4} H_0} \otimes  \mathbb{1}  \; | \mathbb{1}\rangle\rangle_{0}  = \mathbb{1}  \otimes  e^{- \frac{\beta}{4} H_0^T} O^T  e^{-\frac{\beta}{4} H_0^T} \; | \mathbb{1}\rangle\rangle_{0}. \label{equivalentFormsFiniteTemperature}
\end{align}
In particular, for $O = \mathbb{1} $ we have
\begin{equation}
	|\mathbb{1}\rangle\rangle_{\beta}  = e^{- \frac{\beta}{2} H_0}   \otimes  \mathbb{1} \;  | \mathbb{1}\rangle\rangle_{0} =  \mathbb{1}  \otimes e^{-\frac{\beta}{2} H_0^T}  \;  | \mathbb{1}\rangle\rangle_{0}, \label{finiteTemperatureScarInTermsOfInfinte}
\end{equation}
then eq.~\eqref{equivalentFormsFiniteTemperature} can be also written as
\begin{equation}
	|O\rangle\rangle_{\beta} = e^{- \frac{\beta}{4} H_0} O  e^{\frac{\beta}{4} H_0} \otimes  \mathbb{1}  \; | \mathbb{1}\rangle\rangle_{\beta}  = \mathbb{1}  \otimes  e^{- \frac{\beta}{4} H_0^T} O^T  e^{\frac{\beta}{4} H_0^T} \; | \mathbb{1}\rangle\rangle_{\beta}. \label{finiteTemperatureInTermsOfScar}
\end{equation}

Using eq.~\eqref{EPRscar} we have
\begin{equation*}
	\operatorname{Tr}_2 |\mathbb{1}\rangle_{0} \langle \mathbb{1} |_{0} = 	\operatorname{Tr}_2\sum_{n,m}  |n\rangle \langle m | \otimes   |n\rangle  \langle m | =\sum_{n,m}  |n\rangle \langle m | \delta_{nm} = \mathbb{1},
\end{equation*}
where $ \operatorname{Tr}_2 $ is the partial trace with respect to the second layer. Then from eq.~\eqref{finiteTemperatureScarInTermsOfInfinte} we have
\begin{equation}
	\operatorname{Tr}_2 |\mathbb{1}\rangle_{\beta} \langle \mathbb{1} |_{\beta} = \operatorname{Tr}_2 \bigl( ( e^{- \frac{\beta}{2} H_0}   \otimes  \mathbb{1} ) |\mathbb{1}\rangle_{0} \langle \mathbb{1} |_{0}  ( e^{- \frac{\beta}{2} H_0}   \otimes  \mathbb{1} ) \bigr) = e^{- \frac{\beta}{2} H_0} \operatorname{Tr}_2(|\mathbb{1}\rangle_{0} \langle \mathbb{1} |_{0}) e^{- \frac{\beta}{2} H_0} = e^{- \beta H_0}.\label{thermofieldTrace}
\end{equation}
Similarly, the same is true if one traces out the first layer.

\section{S2. Proof of finite-temperature duality}
\label{app:proof}

From eq.~\eqref{relationsBetweenFiniteAndInfinite} and \eqref{equivalentForms} we have
\begin{align*}
	|A O B\rangle\rangle_{\beta} &=  e^{- \frac{\beta}{4} (H_0 \otimes  \mathbb{1} + \mathbb{1} \otimes H_0^T)} |A O B\rangle \rangle_0 \\
	&= e^{- \frac{\beta}{4} (H_0 \otimes  \mathbb{1} + \mathbb{1} \otimes H_0^T)}  A \otimes B^T | O \rangle \rangle_0 = e^{- \frac{\beta}{4} (H_0 \otimes  \mathbb{1} + \mathbb{1} \otimes H_0^T)}  A \otimes B^T e^{ \frac{\beta}{4} (H_0 \otimes  \mathbb{1} + \mathbb{1} \otimes H_0^T)} | O \rangle \rangle_{\beta}.
\end{align*}
for arbitrary operators $A$, $ B$ and $O$. Thus, the dual Hamiltonian takes the form
\begin{equation}
	\mathcal{H}_{\beta} =e^{ -\frac{\beta}{4} (H_0 \otimes  \mathbb{1} + \mathbb{1} \otimes H_0^T)}   \mathcal{H} e^{ \frac{\beta}{4} (H_0 \otimes  \mathbb{1} + \mathbb{1} \otimes H_0^T)}, \label{finiteTemperatureHamiltonianInTermsOfInf}
\end{equation}	
where $	\mathcal{H}$ is obtained form $i \mathcal{L}^{\dagger}$ with dissipator \eqref{finiteTemperature}  via the linear map, which transforms  $A \bullet B \rightarrow  A \otimes B^T  $  for  arbitrary operators $A$ and $ B$, and the Wick rotation $g = i\gamma $. Namely,
\begin{align*}
	\mathcal{H}	&=-H\otimes \mathbb{1}+\mathbb{1}\otimes H^T \nonumber\\
	&+\frac{g}{2}\sum_j  \biggl(  e^{ \frac{\beta}{2}} \left(L_j^\dagger\otimes L_j^T
	-\frac12 L_j^\dagger L_j \otimes\mathbb{1}  -\frac12 \mathbb{1} \otimes  L_j^TL_j^* \right) +  e^{- \frac{\beta}{2}} \left(L_j\otimes L_j^{*}
	-\frac12 L_j L_j^\dagger \otimes\mathbb{1}  -\frac12 \mathbb{1} \otimes  L_j^*L_j^T \right) \biggr).
\end{align*}
From eq.~\eqref{annihilationLike} we have
\begin{align*}
	e^{- \beta H_0} L_j e^{\beta H_0} = e^{\beta} L_j , \qquad e^{- \beta H_0} L_j^{\dagger} e^{\beta H_0} = e^{-\beta} L_j^{\dagger}, \\
	e^{- \beta H_0} L_j L_j^{\dagger} e^{\beta H_0} = L_j L_j^{\dagger}, \quad	e^{- \beta H_0} L_j^{\dagger} L_j e^{\beta H_0} = L_j^{\dagger} L_j.
\end{align*}
Using these relations eq.~\eqref{finiteTemperatureHamiltonianInTermsOfInf} takes the form of eq.~\eqref{finiteTemperatureHamiltonian}.

\section{S3. Classes of dissipative models satisfying condition (\ref{condition for Q})}
\label{app:classes}

A dissipative model is determined by its Hamiltonian $H$ and dissipator $\D$. Condition \eqref{condition for Q} is satisfied by broad classes of dissipative models with  $H$ and $\D$ of varying complexities. We do not attempt to give  an exhaustive classification of such models here; rather we provide a number of examples.

\subsection{Spin $1/2$ models}

Consider a system of $N$  spins $1/2$.

\subsubsection{$\sigma^z$ dissipator}
Consider $\D$ given by the Lindblad operators $L_j=\sigma^z_j,~j=1,2,\dots,N$, $\beta = 0$. Consider further monomials in Pauli matrices,
\begin{equation}\label{V supp 1}
	V^{\alpha_1,\dots,\alpha_n}_{j_1,\dots,\,j_n}=\prod_{l=1}^{n} \sigma_{j_l}^{\alpha_l},\quad \alpha_l\in\{x,y,z\}.
\end{equation}
Then
\begin{equation}\label{condition supp 1}
	\D V^{\alpha_1,\dots,\alpha_n}_{j_1,\dots,\,j_n}=-2\,m\,\gamma V^{\alpha_1,\dots,\alpha_n}_{j_1,\dots,\,j_n},
\end{equation}
where $m$ is the number of $x$- and $y$-Pauli matrices in operator \eqref{V supp 1}. A Hamiltonian that is a linear combination of operators \eqref{V supp 1} with a fixed $m$ satisfies condition \eqref{condition for Q} with $\Gamma=2m\gamma$. In particular,  Hamiltonian \eqref{H exemplary XY} belongs to this class (with $m=2$).

\subsubsection{Isotropic dissipator}
Consider an {\it isotropic dissipator} $\D$ given by the Lindblad operators
\begin{equation}\label{L=sigmaxyz}
	L_{3j-2}= \sigma^x_j,\qquad L_{3j-1}= \sigma^y_j,\qquad L_{3j}= \sigma^z_j,
\end{equation}
where  $j=1,2,\dots,N$, $\beta = 0$.
It is easy to verify that monomials \eqref{V supp 1} are eigenoperators of $\D$,
\begin{equation}\label{condition supp 2}
	\D V^{\alpha_1,\dots,\alpha_n}_{j_1,\dots,\,j_n}=-4\,n\,\gamma V^{\alpha_1,\dots,\alpha_n}_{j_1,\dots,\,j_n}.
\end{equation}
A Hamiltonian that is a linear combination of operators \eqref{V supp 1} with a fixed $n$ satisfies condition \eqref{condition for Q} with $\Gamma=4 n \gamma$.

Let us make a comment on the relevance of the isotropic dissipator to  the exactly solvable Kitaev honeycomb model \cite{Kitaev_2006}. The latter model features a subspace of string-like operators of size $O(N)$ that is closed with respect to commutation with the Hamiltonian  \cite{Gamayun_2022_Out-of-equilibrium}. This space turns out to be also closed  with respect to the action of isotropic dissipator \eqref{L=sigmaxyz}, as can be immediately verified. Thus, augmenting the original Kitaev model with the isotropic dissipator will lead to a dissipative model with an exactly solvable dynamics of $O(N)$ observables. Further, HIoMs of this model will satisfy the condition \eqref{condition for Q}. This is an interesting direction for further research.

\subsubsection{$\sigma^{\pm}$ dissipation and finite temperature}

Let us consider dissipator \eqref{finiteTemperature} with Lindblad operators
\begin{equation}\label{L=sigmamin}
	L_j= \sigma^-_j:=(1/2)(\sigma^x_j-i\sigma^y_j),\qquad  j=1,2,\dots,N.
\end{equation}

These Lindblad operators are neither Hermitian nor unitary, in contrast to those considered previously.

The corresponding dissipator acts on individual Pauli matrices as follows:
\begin{align}
	\D^{\dagger} \sigma^x_j&=-  \frac{\gamma}{2} \cosh \frac{\beta}{2} \sigma^x_j,    \qquad   \D^{\dagger} \sigma^y_j=-  \frac{\gamma}{2} \cosh \frac{\beta}{2} \sigma^y_j,  \nonumber   \\
	\D^{\dagger} \sigma^z_j&=- \gamma\left(\cosh \frac{\beta}{2}\sigma^z_j+\sinh \frac{\beta}{2}\mathds{1} \right). \label{sigma minus dissipator}
\end{align}

Consider the Hamiltonian \eqref{H exemplary XY} with an additional term describing an inhomogeneous  magnetic field in $z$-direction:
\begin{equation}\label{H exemplary XY supp}
	H= \sum_{i<j}\left( w_{ij}^x \,\sigma^x_i \sigma^x_{j}+  w_{ij}^y\, \sigma^y_i \sigma^y_{j} \right) +\sum_{j=1}^N h_j \sigma^z_j.
\end{equation}
Here $h_j$ are real constants. It follows from eq. \eqref{sigma minus dissipator} that
\begin{equation}\label{generalized condition}
	\D^{\dagger} H= -\gamma  \cosh \frac{\beta}{2}  H -\gamma \sinh \frac{\beta}{2} h \cdot \mathds{1},
\end{equation}
where $h=\sum_j h_j$. Due to $ \D^{\dagger}\mathds{1} =0 $ we have
\begin{equation*}
	\D^{\dagger}\left(H + \tanh \frac{\beta}{2} h \cdot \mathds{1} \right) =  -\gamma  \cosh \frac{\beta}{2} \left(H + \tanh \frac{\beta}{2} h \cdot \mathds{1}\right).
\end{equation*}
Thus, condition  \eqref{condition for Q} is met for
\begin{equation*}
	Q = H + \tanh \frac{\beta}{2} h \cdot \mathds{1}.
\end{equation*}

The solution of the GKSL equation in Heisenberg representation reads
\begin{equation}\label{solution with loss}
	H_t=e^{-\gamma \cosh \frac{\beta}{2} t}\, H -  \tanh \frac{\beta}{2}(1-e^{-\gamma \cosh \frac{\beta}{2} t}) h \cdot \mathds{1}.
\end{equation}

Now let us define
\begin{equation}\label{Sz}
	S^z = \frac12 \sum_j \sigma^z_j,
\end{equation}
then from \eqref{sigma minus dissipator} we have
\begin{equation}\label{Q2}
	\D^{\dagger}\left(S^z + \frac{N}{2}\tanh\frac{\beta}2 \cdot {\mathbb 1}\right) = -  \gamma\cosh\frac{\beta}2 \left(S^z + \frac{N}{2}\tanh\frac{\beta}2 \cdot {\mathbb 1}\right).
\end{equation}

Let us calculate
\begin{align*}
		\D^{\dagger}( \sigma^z_j \sigma^z_k) &= (\D^{\dagger} \sigma^z_j)  \sigma^z_k + \sigma^z_j (\D^{\dagger} \sigma^z_k) \\
		&= - \gamma\left(\cosh \frac{\beta}{2}\sigma^z_j+\sinh \frac{\beta}{2}\mathds{1} \right) \sigma^z_k - \gamma  \sigma^z_j \left(\cosh \frac{\beta}{2}\sigma^z_k+\sinh \frac{\beta}{2}\mathds{1} \right)\\
		&= - 2 \gamma \cosh \frac{\beta}{2}\sigma^z_j  \sigma^z_k - \gamma \sinh \frac{\beta}{2} \left(\sigma^z_j +\sigma^z_k\right),
\end{align*}
then from \eqref{Sz} we have
\begin{equation*}
	(S^z)^2 = \frac14 \sum_{j,k} \sigma^z_j  \sigma^z_k  = \frac14 N  \mathds{1} + \frac14 \sum_{j \neq k} \sigma^z_j  \sigma^z_k,
\end{equation*}
hence, we obtain
\begin{equation*}
	\D^{\dagger}\left(	(S^z)^2 - \frac14 N  \mathds{1} \right ) =   - 2 \gamma \cosh \frac{\beta}{2} \left(	(S^z)^2 - \frac14 N  \mathds{1} \right ) - \gamma \sinh \frac{\beta}{2}  (N-1) S^z.
\end{equation*}
Taking into account eq.~\eqref{Q2} we have
\begin{align*}
	\D^{\dagger}&\left(\left(S^z\right)^2 +(N-1)\tanh \frac{\beta}{2}  S^z +\frac{N}{4}\left((N-1) \tanh^2 \frac{\beta}{2} - 1\right)\cdot {\mathbb 1}\right)\\
	&= 2 \, \gamma\cosh\frac{\beta}2 \left(\left(S^z\right)^2 +(N-1)\tanh \frac{\beta}{2}  S^z +\frac{N}{4}\left((N-1) \tanh^2 \frac{\beta}{2} - 1\right)\cdot {\mathbb 1}\right).
\end{align*}

\subsubsection{$\sigma^z\sigma^z$ dissipation}

There was no need to specify any particular geometric structure of interactions in the  Hamiltonians considered above, since each  Lindblad operator acted on a single site. Here we consider Lindblad operators acting on two neighbouring sites of some lattice, and Hamiltonians with nearest-neighbour interactions. Specifically, we consider a linear, square or cubic lattice in $D=1,2,3$ dimensions, respectively. We enumerate the bonds of the lattice by $\upsilon$ and refer to the two distinct sites corresponding to the $\upsilon$'th bond as $i_\upsilon$ and $j_\upsilon$. The dissipator is given by
\begin{equation}\label{L=sigmazsigmaz}
	L_\upsilon= \sigma^z_{i_\upsilon}\sigma^z_{j_\upsilon},
\end{equation}
$\beta = 0$, and the Hamiltonian -- by
\begin{equation}\label{H exemplary XY nearest-neigbour supp}
	H= \sum_\upsilon\left( w_\upsilon^x \,\sigma^x_{i_\upsilon} \sigma^x_{j_\upsilon}+  w_\upsilon^y\, \sigma^y_{i_\upsilon} \sigma^y_{j_\upsilon} \right).
\end{equation}
$\upsilon$ runs over all bonds of the lattice in both formulae above. Note that the Hamiltonian is not necessarily translation-invariant, it is just endowed with the nearest-neighbour structure.

It is easy to see that condition \eqref{condition for Q} is satisfied here:
\begin{equation}\label{condition supp 3}
	\D H = -4\gamma (2 D-1) H.
\end{equation}

It is clear that this example can be generalized to other lattices, next-to-nearest neighbour interactions, as well as Lindblad operators that act on more than two neighboring sites.

\subsection{Generalized Hubbard model}

Here we consider  interacting fermions of two types, $b$ and $c$,  on an arbitrary lattice with $N$ sites, with
\begin{equation}\label{H exemplary Hubbard}
	H= \sum_{i<j}\left( w_{ij}^a \,c^\dagger_i c_j+ w_{ij}^b\,b^\dagger_i b_j\right) + \sum_{i,j} g_{ij} \, c^\dagger_i c_i \, b^\dagger_j b_j
\end{equation}
and dissipator \eqref{finiteTemperature}
\begin{equation}\label{L fermionic}
	L_{2j-1}= b_j, \qquad  L_{2j}= c^\dagger_j c_j, \qquad j =1, 2, \ldots, N.
\end{equation}
The Hamiltonian \eqref{H exemplary Hubbard} generalizes the Hubbard model.

It is easy to see that
\begin{align}\label{action of D on fermions}
	\D\left(n_b + \frac{N}{e^{\beta} -1}\right)&=-\gamma \sinh \frac{\beta}{2} \left(n_b + \frac{N}{e^{\beta} -1} \right), \nonumber \\
	\D(n_c )&=0,
\end{align}
where $n_b = \sum_{j} b_j^{\dagger}b_j$ and $ n_c = \sum_{j} c_j^{\dagger}c_j $.

Eq. \eqref{action of D on fermions} entails that condition \eqref{condition for Q} is satisfied for $Q_b = n_b + \frac{N}{e^{\beta} -1}$ with  $\Gamma_b=\gamma \sinh \frac{\beta}{2}$ and $Q_c = n_c$ with $\Gamma_c = 0$.

Such a dissipator is in detailed balance with $H_0 = n_a$. Remark that it is not hard to generalize this model to the case where $L_{2N+j} = c_j$ are also included in dissipator \eqref{finiteTemperature}, and, moreover, they can be included at a different temperature. Then a multitemperature generalization of \eqref{operatorVectorCorrespondenceFiniteTemperature} with  $\beta H_0 \rightarrow \beta_a n_a + \beta_b n_b$  should be done here to obtain the Hermitian Hamiltonian in the dual picture.

\section{S.4 Generalization of eq. (\ref{dynamics of Q}) }
Here we present a useful generalization of conditions \eqref{HIoM},\eqref{condition for Q} and eq. \eqref{dynamics of Q}. Consider an observable $O$ with the following property: the Heisenberg operator  $O_t|_{\gamma=0}$  in the absence of dissipation
satisfies
\begin{equation}\label{D-condition}
	\D^{\dagger} \left( O_t|_{\gamma=0} \right) =-\Gamma \,\, O_t|_{\gamma=0} .
\end{equation}
One immediately verifies that in this case
\begin{equation}\label{dynamics of O}
	O_t =e^{-\Gamma t} \,  \left( O_t|_{\gamma=0} \right)
\end{equation}
is the solution of the GKSL equation \eqref{Lindblad equation_1} with a finite dissipation. This generalization is important when $\left( O_t|_{\gamma=0} \right)$ is known exactly, e.g. as in Refs. \cite{Prosen_2000_Exact,Lychkovskiy_2021_Closed,Gamayun_2022_Out-of-equilibrium}. In  Ref.~\cite{teretenkov2024exact} eq. \eqref{dynamics of O} was applied to a dissipative $XX$ model. We remark that a similar exponential damping on top of a coherent dynamics has been found theoretically \cite{Caspel_2018_Symmetry-protected,maruyoshi2022conserved} and experimentally \cite{maruyoshi2022conserved} in finite $XXZ$ spin chains with dissipation.

\section{S.5 Operator space fragmentation}

For several models of open quantum systems \cite{dodonov1988evolution, Prosen_2008_Third, Zunkovic_2014_Closed, teretenkov2019dynamics, teretenkov2020dynamics, nosal2020exact, ivanov2022dynamics, linowski2022dissipative, nosal2022higher, Barthel_2022,Zhang_2022, teretenkov2024exact} it was shown that for some operators $R^n$
\begin{equation}\label{closedSubspace}
	\L^\dagger R^n = \sum_m \mathcal{M}_{nm} R^m,
\end{equation}
where $\mathcal{M}_{nm} $ are numbers	and the number of such operators rises polynomially with the system size \footnote{ If one takes as $R_n$ a full matrix basis for a many-body system with finite-dimensional Hilbert space, then $\mathcal{M}_{nm}$ will be just $\L^\dagger$ written in such a basis, but the number of elements of such a basis rises exponentially in the system size.}. 	If we combine the operators $R^n$ in one operator-valued vector $ R $, then eq.~\eqref{closedSubspace} takes the form
\begin{equation}
	\L^\dagger(R) = \mathcal{M} R.
\end{equation}

Moreover, for above-mentioned models the products of operators $R_n$ also satisfy an equation similar to eq.~\eqref{closedSubspace}. If one combines such products in tensors $\otimes^k R$, then it can be expressed
\begin{equation}\label{powerClosedSubspace}
	\L^\dagger( \otimes^k R)  = \mathcal{M}^{(k)} (\otimes^k R),
\end{equation}
where $ \mathcal{M}^{(k)} $ transforms tensors $(\otimes^k R)$ to tensors of the same size. For unitary dynamics eq.~\eqref{powerClosedSubspace} follows directly from  eq.~\eqref{closedSubspace} due to the Leibniz rule for a commutator, which gives $\mathcal{M}^{(k)} = \sum_{j=1}^k \mathcal{M}_j $, where $  \mathcal{M}_j$ acts as $\mathcal{M} $ in the $j$-th subspace and as identity in other tensor multiplicands. As it holds for unitary dynamics one can easily see that eq.~\eqref{powerClosedSubspace}  holds for the Lindbladian with unitary Lindblad operators as well \cite{teretenkov2020dynamics, nosal2020exact, linowski2022dissipative}. In case of Hermitian Linblad operators GKSL dissipators reduce to the sum of  double commutators, which also led to eq.~\ref{powerClosedSubspace} due to the Leibniz rule \cite{ivanov2022dynamics}. In  Ref.~\cite{teretenkov2024exact}  Leibniz-like identity for the commutator of anticommutator  led to the symmetrized analog of eq.~\eqref{powerClosedSubspace} for wide classes of dissipative $XX$ models and Onsager strings as elements of $R$. In Ref.~\cite{nosal2022higher} a generalized  Leibniz-like rule for Lindbladians led to a hierarchical analog of eq.~\eqref{powerClosedSubspace}, where $\L^\dagger( \otimes^k R)$ is a span of  $\otimes^{j} R$ for $j=1, \ldots, k$ rather than just $\otimes^{k} R$.

If the number of operators $R_n$ rises polynomially in the system size, then the number of  $\otimes^k R$ components also does. If $\otimes^k R$ are independent for a large enough system size, then operator space contains exponentially (in the system size) many blocks with closed dynamics, which constitutes operator space fragmentation \cite{Essler_2020_Integrability}. Duality \eqref{operatorVectorCorrespondenceFiniteTemperature} brings such operator spaces into Hilbert spaces
\begin{equation}
	\operatorname{span}   |\otimes^k R \rangle \rangle_{\beta} := \operatorname{span}  | R^{n_1} \cdot \ldots \cdot R^{n_k}\rangle \rangle_{\beta}
\end{equation}
invariant under  dynamics with  dual Hamiltonian \eqref{finiteTemperatureHamiltonian} inducing  Hilbert space fragmentation for such a Hamiltonian. Let us emphasize that the structure of fragmented operator spaces induced by operators products $ R^{n_1} \cdot \ldots \cdot R^{n_k}$ or  symmetrized products is more evident than the structure of the dual fragmented Hilbert spaces.

\section{S.6 First few states in the tower of scars}

Here we provide explicit expressions for two scars of the model \eqref{dual sigma pm example} dual to eigen-HIOMS \eqref{HIoMs exemplary}. According to eq. \eqref{additional scars}, these (unnormalized) scar eigenstates and eigenenergies  read
\begin{align}
	|Q_1\rangle\rangle_\beta & = |S^z\rangle\rangle_\beta + \frac{N}{2}\tanh\frac{\beta}2 \, |{\mathbb 1}\rangle\rangle_\beta,            & E_1 & = - g \cosh\frac{\beta}2,  \nonumber \\
	|Q_2\rangle\rangle_\beta & =| \left(S^z\right)^2\rangle\rangle_\beta +(N-1)\tanh \frac{\beta}{2}  |S^z\rangle\rangle_\beta 	+\frac{N}{4}\left((N-1) \tanh^2 \frac{\beta}{2} - 1\right)\, | {\mathbb 1}\rangle\rangle_\beta,        &
	E_2 & = - 2 \, g\cosh\frac{\beta}2, \label{tower of scars}
\end{align}
with $|{\mathbb 1}\rangle\rangle_\beta$ given by eq. \eqref{thermofield double state} and
\begin{align}
	|S^z \rangle\rangle_\beta & =\frac12 \sum_{j=1}^N   \left(
\Big(e^{-\frac{\beta}{4}}|\uparrow\,\uparrow\rangle_j-e^{\frac{\beta}{4}}|\downarrow\,\downarrow\rangle_j\Big)
\bigotimes_{\substack{l=1\\ l\neq j}}^N  \Big(e^{-\frac{\beta}{4}}|\uparrow\,\uparrow\rangle_l+e^{\frac{\beta}{4}}|\downarrow\,\downarrow\rangle_l\Big)
\right), \nonumber \\
	|\left(S^z\right)^2 \rangle\rangle_\beta & =\frac{N}4\,|{\mathbb 1}\rangle\rangle_\beta +\frac14 \sum_{i<j}  \left(
\Big(e^{-\frac{\beta}{4}}|\uparrow\,\uparrow\rangle_i-e^{\frac{\beta}{4}}|\downarrow\,\downarrow\rangle_i\Big)\otimes
\Big(e^{-\frac{\beta}{4}}|\uparrow\,\uparrow\rangle_j-e^{\frac{\beta}{4}}|\downarrow\,\downarrow\rangle_j\Big)
\bigotimes_{\substack{l=1\\ l\neq i,j}}^N \Big(e^{-\frac{\beta}{4}}|\uparrow\,\uparrow\rangle_l+e^{\frac{\beta}{4}}|\downarrow\,\downarrow\rangle_l\Big)
\right).
\end{align}
As noted in the main text, an additional scar of the model \eqref{dual sigma pm example} can be obtained  from the single-layer Hamiltonian $H$. This scar explicitely reads
\begin{align}
	|H \rangle\rangle_\beta & =2\sum_{i<j} w_{ij} \left(
\Big(|\uparrow\,\downarrow\rangle_i\otimes |\downarrow\,\uparrow\rangle_j+ |\downarrow\,\uparrow\rangle_i \otimes |\uparrow\,\downarrow\rangle_j\Big)
\bigotimes_{\substack{l=1\\ l\neq i,j}}^N \Big(e^{-\frac{\beta}{4}}|\uparrow\,\uparrow\rangle_l+e^{\frac{\beta}{4}}|\downarrow\,\downarrow\rangle_l\Big)
\right).
\end{align}

Note that these scars have nonzero entanglement for the transverse bipartition, in contrast to the thermofield scar~$| {\mathbb 1}\rangle\rangle_\beta$.

\section{S.7 A different type of scar states in bilayer systems}

Consider a bilayer system with the Hamiltonian

\begin{align}\label{bilayer general}
	\mathcal{H} =& H_1+H_2+ \sum_{i=1}^{N_1} \sum_{j=1}^{N_2} g_{ij} \bigl(\sigma^x_i \tau^x_j - \sigma^y_i \tau^y_j \bigr),	
\end{align}
where $H_1$ ($H_2$) is a Hamiltonian of $N_1$ ($N_2$) spins $1/2$ constituting the first (second) layer, and the third term in $\mathcal{H}$ is the interlayer interaction. The Hamiltonians $H_1$ and $H_2$ conserve the $z$-projections $S^z_1=\sum_{i=1}^{N_1}\sigma^z_i$ and  $S^z_2=\sum_{j=1}^{N_2}\sigma^z_j$  of spins of respective layers. Apart from this,  $H_1$ and $H_2$ are arbitrary (in particular, can contain a different number of spins), and therefore the closed system $\mathcal{H}$ is not covered by the duality discussed in the main text. Still, the states
\begin{equation}
|P_1\rangle=|\underbrace{\uparrow\uparrow\dots\uparrow}_{1\mathrm{st~  layer}}\rangle\otimes|\underbrace{\downarrow\downarrow\dots\downarrow}_{2\mathrm{nd~  layer}}\rangle, \qquad |P_2\rangle=|\underbrace{\downarrow\downarrow\dots\downarrow}_{1\mathrm{st~  layer}}\rangle\otimes|\underbrace{\uparrow\uparrow\dots\uparrow}_{2\mathrm{nd~  layer}}\rangle
\end{equation}
are eigenstates of $\mathcal{H}$. Indeed, each of this states is a maximal(minimal)-eigenvalue eigenstate of $S^z_1$ and $S^z_2$ and hence of  $H_1$ and $H_2$. Further, each of this states is a zero-eigenvalue eigenstate of each term of the form $(\sigma^x_i \tau^x_j - \sigma^y_i \tau^y_j )$, which completes the proof.

In a special case of $\mathcal{H}$ dual to an open system according to eq. \eqref{finiteTemperatureHamiltonian}, the eigenstates $|P_1\rangle$ and $|P_2\rangle$ are dual to non-Hermitian eigen operators of the Lindbladian
\begin{equation}
P_1 = \prod_{j=1}^N \sigma_j^+, \qquad P_2= \prod_{j=1}^N \sigma_j^-.
\end{equation}

\end{document}